**Article Title:** Software Startup Practices – Software Development in Startups through the Lens of the Essence Theory of Software Engineering

**Authors:** Kai-Kristian Kemell, Ville Ravaska, Anh Nguyen-Duc, and Pekka Abrahamsson

**Year:** 2020

# Software Startup Practices – Software Development in Startups through the Lens of the Essence Theory of Software Engineering

Kai-Kristian Kemell[1][0000-0002-0225-4560] Ville Ravaska[1], Anh Nguyen-Duc[2][1111-2222-3333-4444] and Pekka Abrahamsson[1][0000-0002-4360-2226]

[1] University of Jyväskylä, Jyväskylä 40014, Finland
[2] University of Southeast Norway, Norway
kai-kristian.o.kemell@jyu.fi

**Abstract.** Software startups continue to be important drivers of economy globally. As the initial investment required to found a new software company becomes smaller and smaller resulting from technological advances such as cloud technology, increasing numbers of new software startups are born. Typically, the main argument for studying software startups is that they differ from mature software organizations in various ways, thus making the findings of many existing studies not directly applicable to them. How, exactly, software startups really differ from other types of software organizations as an on-going debate. In this paper, we seek to better understand how software startups differ from mature software organizations in terms of development practices. Past studies have primarily studied method use, and in comparison, we take on a more atomic approach by focusing on practices. Utilizing the Essence Theory of Software Engineering as a framework, we split these practices into categories for analysis while simultaneously evaluating the suitability of the theory for the context of software startups. Based on the results, we propose changes to the Essence Theory of Software Engineering for it to better fit the startup context.

## 1 Introduction

Software startups continue to be important drivers of economy globally. As the initial investment required to found a new software company becomes smaller and smaller resulting from technological advances such as cloud technology, increasing numbers of new software startups are born. While most startups fail [3], much like any other type of new company [11], some go on to become mature, established software organizations, or even multinational tech giants.

Typically, the main argument for studying software startups is that they differ from mature software organizations in various ways, thus making the findings of many existing studies not directly applicable to them. This is a result of there still being no accurate definition for what a startup is [18][20]. Various characteristics such as time pressure or resource scarcity are attributed to startups to differentiate them from mature companies [18], but academically drawing an exact line has been a challenge in the area [11]. Nonetheless, this has also been true in relation to the way software startups develop software. Existing studies have sought to understand how software startups develop software compared to mature software organizations and how they differ in terms of software development.

For example, Paternoster et al. [18] conducted a more general, large-scale study aiming to understand how software startups develop software. They noted that software startups operate mostly using Agile methods or ad hoc methods. Specific facets of software development in software startups, such as prototyping [16] have also been studied. However, studies focusing on Software Engineering (SE) practices in software startups are still scarce, and studies into software development in software startups in general are still needed [20]. Some high-profile practices utilized by startups such as the Five Whys are commonly discussed in e.g. startup education, but systematic studies into the topic are lacking.

Thus, to better understand how software development in software startups might differ from software development in more mature software organizations, we study practices in this paper. Specifically, we seek to understand what practices are commonly used by software startups (**RQ1**). In addition, we approach this topic through the lens of the Essence Theory of Software Engineering and seek to understand how this theory fits into the context of software startups. To this end, we study how the seven alphas of the theory – which we discuss in the second section – are suited for the context of software startups, and whether other alphas would be needed to make the theory better suited for the startup context (**RQ2**).

## 2 Background – Software Startups, Software Development Practices, and the Essence Theory of Software Engineering

This section is split into three subsections. First, we briefly discuss software startups in relation to software development. Then, we define what we refer to with software development practices.

### 2.1 Software Development in Software Startups

Typically, software startups do not strictly follow any formal software development method [18]. Instead, they combine practices from different methods that suit their needs at the moment or simply use ad hoc practices [15].

As the aim of this study is to uncover software development practices universal to (most) software startups, a notable study is that of Dande et al. [5]. Dande et al. [5] studied software startups in Finland and Switzerland and devised a list of 63 practices commonly utilized by software startups. However, these practices are not solely software development ones but also include practices related to customers and business. Kamulegeya et al. [9] studied these practices and reported that they seemed to apply in the Ugandan startup context as well, providing further validity to the practices. However, they do add that while they consider most of the listed practices to be universal, some might vary in different locations or cultures.

Other studies focusing on practices have not aimed to create such extensive lists of practices but have nonetheless studied software startup practices in different contexts. Klotins, Unterkalmsteiner, and Gorschek [13], for example, created a framework for categorizing software startup practices that differs from the one proposed by Dande et al. [5]. Giardino et al. [7] propose the Greenfield Startup Model to explain software development in early-stage software startups. In the process, they uncovered various practices that supplement and confirm the findings of Dande et al. [5]. Paternoster et al. [18] in their study on how software startups develop software discuss having found 213 practices, which, however, were not listed in their paper. Nonetheless, their findings to lend support to those of Dande et al. [5].

### 2.2 Software Development Practices

Jacobson et al. [8] suggest that a set of practices is what forms a method in the context of SE. Methods, according them, describe ways-of-working, i.e. how work should be carried out. Whether or not formal SE method is employed, a way-of-working exists in the form of an informal method that could nonetheless be described if necessary [8]. In this logic, practices are more atomic units that describe how work is carried out.

Historically, in academic literature and particularly in Information Systems, the construct *technique* has been used for the same purpose in the context of method engineering [19]. Tolvanen [19] defines a technique to be a set of steps and rules that define how a representation of information system is derived and handled using conceptual structure and related notation. A tool, in this context, refers to a computer-based application supporting the use of a technique.

## 2.3 The Essence Theory of Software Engineering

The Essence Theory of Software Engineering [8] provides a way of describing methods and practices. It consists of a notational language, which is used to achieve this, as well as a so-called kernel, which includes building blocks which one can use as a basis for constructing methods. The kernel, its authors argue [8], contains basic elements that are universal in any SE project.

The Essence kernel contains three types of objects: alphas (i.e. things to work with), activities (i.e. things to do), and competencies (skills required to carry out the work). In this study, we focus on the alphas in the context of software startups. The seven Essence alphas are as follows: (1) Stakeholders, (2) Opportunity, (3) Requirements, (4) Software System, (5) Team, (6) Way of Working, and (7) Work. These alphas are split into three areas of concern. The first two belong in the customer area of concern, numbers three and four in the solution area of concern, and the last three in the endeavor area of concern. [8] Furthermore, each alpha has alpha states used to track progress on the alpha. [8]

The authors of Essence posit [8] that these are the essential elements that are present in every SE project. Every project, then, has its own unique context, which most likely contains more things to work with, but those are not universal to every project. In order to reap the most benefits out of Essence, its users would then extend this basic kernel with the Essence language to include these unique features of their particular project or company to describe their method(s) with it.

In this paper, the role of Essence is two-fold. First, it serves as a framework for analyzing our data. We utilize the alphas to sort the software startup practices we discover into categories. Secondly, in the process of doing so, we study whether all the uncovered practices fit into these seven alphas. I.e., do the alphas also present all the essential elements of software development in software startups?

## 3 Study Design

The aim of this study is to better understand how software startups develop software by focusing on practices. We approach this topic using a qualitative approach. This study utilizes both empirical data and the list of startup practices presented by Dande et al. [5] who list 63 software startup practices. With the data, we seek to both further validate this list of 63 practices, and to potentially uncover new ones.

### 3.1 Data Collection

The empirical data for this study was collected by means of a multiple case study (n=13) (Table 1). We utilized a qualitative approach and data from the cases were collected with thematic interviews. We chose a thematic approach because most software startups develop software ad hoc [15][18] and use different terminology than the academia, which presents challenges for a structured interview approach.

**Table 1.** Cases.

| Case | Employees | Company Domain | Respondents |
|------|-----------|----------------|-------------|
| 1 | 6 | Software/ Hardware | 1 |
| 2 | 5 | Software | 3 |
| 3 | 3 | Software / Hardware | 2 |
| 4 | 5 | Software | 1 |
| 5 | 7 | Software / Consulting | 1 |
| 6 | 3 | Software / Hardware | 1 |
| 7 | 8 | Software | 1 |
| 8 | 12 | Software | 1 |

| 9 | 6 | Software | 1 |
| 10 | 5 | Software | 1 |
| 11 | 85 | Software / Hardware | 1 |
| 12 | 5 | Software / Hardware | 1 |
| 13 | 6 | Software | 1 |

The themes for the interviews were the seven Essence alphas (Section 2.3). The approach was semi-structured in nature, with some questions prepared beforehand, and with more questions asked based on the answers of each respondent. Aside from these seven themes, the respondents were asked for background information on their startup and their role in it, as well as background information on the startup's journey thus far (e.g. idea initiation, current state, challenges faced etc.).

The interviews were conducted F2F. The audio was recorded, and the recordings were transcribed for analysis.

### 3.2 Data Analysis

The analysis of the empirical material in this thesis was conducted following the thematic synthesis guidelines of Cruzes and Dyba [4]. The material was first transcribed. The external interviews were transcribed by the original researchers. Second the material was read through thoroughly and some notes were made from the material. After this step, the interviews were coded, and the codes were categorized under different themes. The original themes were then fitted to the theoretical framework of this thesis and those themes that did not fit under the categories of the framework were analysed by creating additions to the original theoretical framework.

Practices that were discussed by two or more of the case startups were considered prevalent enough to be included into the list of practices. Once the empirical data had been analyzed and new practices had been formulated, we took the list of 63 software startup practices of Dande et al. [5] and these new practices and inserted them into the framework of the Essence Theory of Software Engineering [8] and its alphas. I.e., we categorized each practice, if possible, under one of these alphas. This process was carried out by one author, after which the list was reviewed by two other authors and modified to form a mutual consensus.

## 4 Results

This section is divided into 9 subsections. In the first one, we present the new practices we uncovered through the case study. In the next seven, we go over the results in relation to each Essence alpha, discussing the practices found in each category. In the ninth and final one, we discuss practices that did not fit under any of these alphas.

In the interest of space in this paper, the clarifying descriptions for the 63 practices of Dande et al. [5] have not been included in the tables in this section. Such descriptions have, on the other hand, been added for any novel practices proposed by us. Each practice has an identified (Pn), where practices P64 and up are practices based on the empirical data and practices P63 and below are from Dande et al. (2014).

### 4.1 New Practices

Based on the data, we propose 13 new practices (Table 2) that were not present in the list of Dande et al. [5]. These practices were clearly discussed by at least two case startups. Other new practices were also uncovered but discussed by only one case startup. These practices were not considered 'common' based on this set of data. However, as this approach was qualitative, we nonetheless list these practices in the last subsection of this section for future validation.

**Table 2.** New practices based on our data.

| ID | Practice | Description |
|---|---|---|
| P64 | Study subjects that support the startup | Studying while working on a startup gains competence in the team without growing in personnel. |
| P65 | Attend startup events | Startup events provide opportunity for feedback from experts and allows you to meet potential investors. |
| P66 | Create an MVP early on | MVP helps you to focus on the most important features in the beginning. |
| P67 | Test features with customers | Testing features with real customers gets you the best feedback. |
| P68 | Get advisors | Experienced professionals or investors can help startup to grow in advisor or mentor role. |
| P69 | Use efficient tools to plan your business model | Business model canvas, pitch deck etc. help you to focus your business idea and are easy to change if needed. |
| P70 | Test different tools | Start with tools team is familiar with and test different ones to find those that work the best for you. |
| P71 | Conduct market research | Research the markets and competitors to focus your idea and to find your unique value proposition. |
| P72 | Have frequent meetings with the whole team | Use meetings to organize and plan your work at least once a week. |
| P73 | Avoid strict roles | Let the team co-operate in all of the tasks. |
| P74 | Create a prototype | Create prototype to validate your product or features. |
| P75 | Use efficient communication tools | Use tools that allow natural communication inside the team when not working in the same space. |
| P76 | Prioritize features | Choose which features are needed now and plan others for future releases. |

## 4.2 Opportunity

The opportunity alpha is in the intersection between business and software development. In the Essence theory, it is about understanding the needs the system is to fulfill and is in the customer area of concern. Practices for this alpha are presented in Table 3 below. No new practices for this category were found in the data.

The case startups were highly focused on understanding their customers and fulfilling the needs of the customer (segments). This is in line with the idea of software startups being product-oriented and customer-focused. On the other hand, the lack of support for P4 makes it seem that these startups were more focused on fulfilling the needs they had uncovered rather than understanding why these needs were important.

Focusing on the system and the needs it was intended to fulfill was considered important from the point of view of competition as well. Focusing on one's unique value proposition is conventionally considered an important strategy for differentiating from one's competitors.

**Table 3.** Practices for the Opportunity alpha.

| ID | Practice | Cases Supporting |
|---|---|---|
| P1 | Focus your product | 1,2,6,7,8,9,11,12,13 |
| P2 | Find your value proposition and stick to it on all levels | 9,13 |
| P4 | Focus on goals, whys | 9 |
| P18 | Validate that your product sells | 1,2,4,5,7,8,11 |
| P20 | Form deep relations with the first customers to really understand their needs | 1,6,9,11,13 |
| P33 | In the development of customer solutions, find a unique value proposition in your way of acting | 1,2,3,5,6,8,9 |
| P34 | Follow communities | 1,2 |

## 4.3 Stakeholders

Four practices were categorized under the stakeholder alpha (Table 4), which is another alpha in the customer area of concern in Essence. For startups, most notable stakeholders are typically investors and customers or users. Investors provide funding and potentially other resources, while customers are the ones whose needs are to be fulfilled by the system. In addition, nearly half of the case startups discussed the importance of their advisors as another group of stakeholders (P68).

**Table 4.** Practices for the Stakeholders alpha.

| ID | Practice | Cases Supporting |
|---|---|---|
| P24 | Keep customer communications simple and natural | 6 |
| P32 | Showing alternatives is the highest proof of expertise | - |
| P35 | Share ideas and get more back | 1,2 |
| P68 | Get advisors | 1,4,5,6,8,9 |

Especially early-stage startups tend to rely on advisors who may also be or later become investors in some cases. For example, startup ecosystems tend to foster advisor relationships in various ways. Startups working in incubators are likely to receive guidance from various experts, who may eventually form an even closer advisor relationship with the startup. Advisors can provide startups with capabilities they are lacking, e.g. accounting, as well as help them expand their contact networks.

The practice of sharing ideas to hone them and to get feedback was also discussed by same case startups. While in some cases companies may be reluctant to share their ideas in fears of having them stolen, none of the case startups indicated this type of concerns. To this end, advisors can also provide feedback if a startup is afraid of revealing their ideas to potential investors due to such concerns.

## 4.4 Requirements

Requirements help provide scope for the work being done on the system. The requirements alpha is in the solution area of concern of the Essence kernel. Four new practices were uncovered in this category and most existing practices in this category were well-supported by the cases (Table 5).

However, one of the practices of Dande et al. [5], P3, was in conflict of what some of the case startups stated. While P3 has it that startups should present their product as facilitating rather than competing to their competitors, in some cases the startups do seek to compete in various ways. The facilitating approach is certainly a valid approach as well, if mutually exclusive.

The requirements alpha was closely related to the stakeholders alpha, as the customer-focus through various validation activities made the requirements highly determined by the uncovered customer needs (P10). In the case startups, prototypes were typically used to do carry out validation (P67, P74). While a startup should be open to new features and needs (P51), they should be prioritized (P76) to create a clear core product (P52, P53).

**Table 5.** Practices for the Requirements alpha.

| ID | Practice | Cases supporting | Cases conflicting |
|---|---|---|---|

| ID | Practice | Cases supporting | Cases conflicting |
|----|----------|------------------|-------------------|
| P3 | Present the product as facilitating rather than competing to the competitors | - | 1,2,6 |
| P5 | Use proven UX methods | 12 | - |
| P10 | Design and conduct experiments to find out about user preferences | 1,2,4,6,9,12,13 | - |
| P21 | Use planning tools that really show value provided to customers | 2 | - |
| P51 | Anything goes in product planning | 1,2,11 | - |
| P52 | To minimize problems with changes and variations develop a very focused concept | 1,2,3,4,5,6,7,12,13 | - |
| P53 | Develop only what is needed now | 1,2,3,12 | - |
| P66 | Create an MVP in the beginning | 1,2,4,13 | - |
| P67 | Test features with customers | 1,3,4,5,6,7,8,9,11 | - |
| P74 | Create prototype | 1,2,3,4,5,6,9,12 | - |
| P76 | Prioritize features | 1,2,3,9,11 | - |

## 4.5 Software System

The software system alpha is focused on the product itself, i.e. the system; software or hardware. The software system alpha is in the solution area of concern of the Essence kernel. Some of the previously proposed practices were largely prevalent in the cases while some received little support from our data. More technical practices (P23, P54, P57) would have required a more technical focus from the interviews. No new practices were proposed for this category. The practices for this category are in Table 6.

Out of the practices of this category, only P7 had some conflicts in the data. This practice is largely B2C focused, whereas a B2B startup might understandably focus on tailoring its system especially for larger customers. However, it is perhaps worth aiming for a modular product where such manual tailoring is not needed. Most case startups nonetheless adhered to this practice.

Overall, these practices further underlined that startups should have a clear focus in their development. For example, they should focus on a limited number of platforms, possibly only one initially (P8). Additionally, startups are conventionally seen as agile and their systems as prone to changes based on feedback. Indeed, these practices support the idea that the system should be developed with modifications in mind (P60). Features should be easily added (P55) or removed (P54) when necessary.

**Table 6.** Practices for the Software System alpha.

| ID | Practice | Cases supporting | Cases conflicting |
|----|----------|------------------|-------------------|
| P7 | Have a single product, no per customer variants | 1,2,3,5,7,8,11,12 | 6,13 |
| P8 | Restrict the number of platforms that your product works on | 1,2,3,4,7,12 | - |
| P14 | Anyone can release and stop release | 2 | - |
| P23 | Adapt your release cycles to the culture of your users | - | - |
| P54 | Make features easy to remove | - | - |
| P55 | Use extendable product architecture | 1,2,3,9,11 | - |
| P57 | Bughunt | - | - |
| P58 | Test APIs automatically, UIs manually | 2,13 | - |
| P59 | Use generic, non-proprietary technologies | 2,7 | - |
| P60 | Create a solid platform | 3,8,9,11 | - |

## 4.6 Work

Work in the context of Essence refers to the work tasks required to produce the system. It is under the endeavor area of concern in the Essence kernel. For software startups, this also involves business model development. How the work is carried out from the point of view of e.g. methods, belongs into the way of working category, on the other hand. Few existing practices were considered to belong into this category and no new practices for this category were found (Table 7).

**Table 7.** Practices for the Work alpha.

| ID | Practice | Cases supporting |
|---|---|---|
| P44 | Tailored gates and done criteria | 8 |
| P48 | Fail fast, stop and fix | 1 |
| P62 | Use the most efficient programming languages and platforms | 2,3,7 |

While P48 is arguably closely related to prototyping and validation activities which were extensively discussed by the respondents, it was seldom discussed directly. On the other hand, P62 was discussed in relation to system architecture. Efficiency in this case was considered subjectively: the developers focused on languages and platforms they had prior experience with and could thus start working the fastest with.

## 4.7 Team

The team comprises the individuals working on the startup, the founders or owners and the employees or unpaid ones. It is under the endeavor area of concern in the Essence kernel. The team sizes for the case startups are in Table 1 in Section 3. One new practice (P64) was added into this category based on the data (Table 8).

**Table 8.** Practices for the Team alpha.

| ID | Practice | Cases supporting | Cases conflicting |
|---|---|---|---|
| P26 | Flat organization | 1,2,3,5,9 | - |
| P27 | Consider career expectations of good people | 4,9 | - |
| P28 | Don't grow in personnel | 1,2,3,12 | - |
| P29 | Bind key people | 2,3,6,7 | - |
| P36 | Small co-located teams | 1,2,3,4,5,6 | 12 |
| P37 | Have multi-skilled developers | 1,2,3,12 | - |
| P38 | Keep teams stable in growth mode | 1,2,3,4,6,7,13 | 9 |
| P40 | Sharing competence in team | 4,5 | - |
| P41 | Start with competence focus and expand as needed | 1,2,3,4,6,8,9,13 | - |
| P42 | Start with small experienced team and expand as needed | 1,2,3,4,7,8,12,13 | 1,2,3 |
| P64 | Study skills and topics that support your startup | 1,2,3,4,8,9 | - |

The most commonly discussed practices were P41 and P42. The initial team of a startup is important as it needs to have the required competencies to carry out the work (P41). To this end, an experienced team may be required (P42). Some of the case conflicted with P42. However, this did not mean that the startups did not want and experienced team. Rather, they simply did not have one due to being founded by a group of students with little prior experience.

If the team is lacking competencies and expanding the team is not possible or feasible in a given situation, the existing team members may be have to learn new skills instead (P64). This also ties to P37, as the small team sizes often result in a single employee having to take on various different tasks. A developer is often involved in business decisions as well, especially in early-stage startups.

Flat organization structures (P26) are commonly attributed to startups and this was the case in our data as well. Involving employees in decision-making may also serve to better bind them to the startup (P29). With a small, focused team, staff turnover can be damaging (P38).

## 4.8   Way of Working

Way of Working refers to how the work is carried out, including practices, tools, processes, and methods [8]. It is under the endeavor area of concern in the Essence kernel. Most previously proposed practices were supported by our data in this category. Four new practices were proposed for this category (Table 9).

Most case startups discussed having taken some existing agile practices and tailoring them rather than using them by the book (P47). While this ties to P72 in that frequent team meetings are common in agile development, it gained enough emphasis to be its own separate practice. On the other hand, the use of formal methods (P46) was not discussed by any of the startups, with the teams seemingly utilizing various individual practices to compose their own.

**Table 9.** Practices for the Way of Working alpha.

| ID | Practice | Cases supporting | Cases conflicting |
|----|----------|------------------|-------------------|
| P9 | Use enabling specifications | 1,2,3 | - |
| P15 | Create the development culture before processes | 1,8,11 | - |
| P39 | Let teams self-select | 1,2,3,5,8 | - |
| P43 | Have different processes for different goals | - | - |
| P45 | Time process improvements right | 3 | - |
| P46 | Find the overall development approach that fits your company and its business | - | - |
| P47 | Tailor common agile practices for your culture and needs | 1,2,3,4,6,7,8,13 | - |
| P49 | Move fast and break things | 4,7 | - |
| P50 | Forget Software Engineering | 1 | - |
| P61 | Choose scalable technologies | 2,3,9,11 | - |
| P63 | Start with familiar technologies and processes | 1,2,3,7 | - |
| P70 | Test different tools | 1,3 | - |
| P72 | Have frequent meetings with the whole team | 1,2,3,4,5,8,12 | - |
| P73 | Don't have strict roles | 1,2,3 | 9 |
| P75 | Use efficient communication tools | 2,3,5 | - |

To this end, communication in general is an important part of agile development, and arguably development in general. The case startups frequently discussed the importance of tools in facilitating communications (P75). While shared physical workspaces can reduce the need for such, their importance is further highlighted when working remotely. Moreover, an early-stage startup may not have a dedicated physical workspace, or its members may have regular jobs alongside their startup activities, resulting in erratic work hours in the startup. This makes communication tools such as instant messaging services important.

Self-organizing teams are recommended in agile development and this is also arguably common for startups (P39, P73).

## 4.9 Other Practices Unsuited for Existing Essence Alphas

Not all of the practices we propose, or the ones proposed by Dande et al. [5], fit under any of the existing Essence alphas. These were practices related to the business aspect of software startups, such as marketing, business model development, or funding. Whereas Essence focuses on SE in mature software organizations, the business aspect in software startups is closely intertwined with software development. For example, the needs of the customers or the customers in general, may not be clear to a software startup, which results in the requirements evolving over time.

**Table 10.** Practices not applicable to any existing Essence alpha.

| ID | Practice | Case supporting | Case conflicting |
|----|----------|-----------------|------------------|
| P6 | Do something spectacular | - | - |
| P11 | Use tools to collect data about user behavior | 1,2,7 | - |
| P12 | Make your idea into a product | 1,2,3,4,5,6,7,8,12,13 | 11 |
| P13 | Outsource your growth | 5,9,11,12,13 | 3 |
| P16 | Get venture capital and push your product | 1,2,4,5,8,9 | 3 |
| P17 | Fund it yourself | 1,2,3,7,9 | - |
| P19 | Focus early on those people who will give you income in the long run | 5,6,7,8,11,13 | - |
| P22 | Start locally grow globally | 1,2,3,6,7,8,9,13 | - |
| P25 | Help customers create a great showcase for you with support | 1,6,8,9 | - |
| P30 | Form partnerships and bonds with other startups | 1,3,4,5,13 | - |
| P31 | Make your own strength as a "brand" | 8 | - |
| P56 | Only use reliable metrics | 5,6,7 | - |
| P65 | Attend startup events | 1,2,3,4,8 | - |
| P69 | Use efficient tools to plan your business model | 1,2,3 | - |
| P71 | Conduct market research | 1,2,6,12 | - |

Practices P6, P11, P25, P31, and P71 concern marketing activities. For example, P25 is about getting a few initial customers who are particularly interested in the system and who can then be used as reference customers in marketing, or who themselves can market the product. P6 and P31 are more general marketing practices. These types of activities are difficult to incorporate into any existing Essence alpha. While marketing is a customer related activity and thus could be linked to stakeholders, the existing stakeholder alpha focuses on clearly identified and involved stakeholders such as the organization commissioning a project, as opposed to obtaining new customers (stakeholders).

P16 and P17 are related to funding. Funding or simply available cash to burn is something that is constantly tracked in a startup, much like the alphas are tracked in Essence. No existing alpha supports funding with clear emphasis. Some of the alpha states of the Work alpha include mentions of securing sufficient funding, but this process is seldom so straightforward in a startup.

The remaining practices in this category are related to overall business model development and business planning. For example, P13 suggests that outsourcing some part of the business can help the startup focus on the core product, and P22 suggests a strategy for rapid and high growth. P30, on the other hand, could be filed under the Stakeholders alpha, but doing so might not place sufficient emphasis on the strategic importance of such decisions from a business point of view.

As we do not formally develop new alphas in this paper, we leave the proposals related to these observations for the following discussion section.

## 5   Discussion

The theoretical contribution for this study is related to the Essence Theory of Software Engineering, which we have used as a theoretical framework in this study. Essence is intended to be used in any SE endeavor. Its so-called kernel, its authors argue [8], contains the elements present in every SE endeavor. This kernel acts as a set of building blocks that can then be extended using the Essence language to describe methods.

In this paper, we looked at Essence from the point of software startups. Much effort has gone into understanding how exactly software startups differ from other types of organizations, including how they develop software [18][20], as was the focus here. As startups are considered to differ from other types of organizations in various ways, existing theories focused on traditional companies may not fully work in their context.

In software startups, the business aspect is deeply intertwined with software development [13]. In fact, Klotins et al. [12] argue that software startups most of the time fail due to business issues that in fact originate from SE processes. This supports the idea that software development and business are difficult to separate in software startups and should be presented in the same framework. Essence would have to be modified to achieve this.

The Essence Theory of Software Engineering is focused on conventional software project work and does not take business aspects into account, as they are not similarly relevant. A commissioned project has clear requirements, and a company developing a new version of an existing product already has customers that can be used to elicit requirements. Software startups, on the other hand, work with unclear requirements and have to actively seek to validate their idea as they go. This can also be the case with new product development in mature companies seeking to expand into new market. Moreover, mature organizations are typically larger and have clearer roles. A developer typically simply develops, generally unconcerned about the business aspects of the company such as marketing, whereas in software startups developers typically are involved in business development as well, as was seen in many of the case startups of this study, too.

In categorizing the practices proposed in this paper, as well as those proposed by Dande et al. [5], we found some of the practices unsuited for any of the existing alpha categories. Namely, the business-related practices were not well-suited for the seven alphas. While the opportunity and stakeholder alphas cover some user-related business activities, many practices are difficult to incorporate into these alphas.

We thus propose that the Essence kernel should be expanded for software startup use. The following changes should be done:

- A fourth area of concern to encompass the business aspects should be added.
- New alphas for the business area of concern should be added.

Alphas are things to work with and while using Essence one tracks progress on the alphas, each of which is split into alpha states to aid in this process. Thus, the new alphas should be measurable in some fashion. We propose three new potential alphas: (1) funding, (2) business model, and (3) marketing.

First, funding is pivotal for any startup [2]. It can be quantitatively measured and various widely used startup metrics support measuring it. It makes for a straightforward alpha in this regard. It is likely that progress on this alpha is not permanent at all but fluctuates based on the situation. Whereas a system steadily progresses towards being more and more complete, funding is spent and thus its status can fluctuate continuously as old funding is used and new funding is obtained.

Secondly, business model development is at the core of a startup [14]. Indeed, one widely used definition for what is a startup posits that a startup is a "temporary organization designed to look for a business model that is repeatable and scalable" [1]. A startup spends notable amounts of resources seeking to create a functioning business model, and many of the practices discussed in this paper as well play a key role in that process. E.g., various validation activities focused on the users or customers that belong under the stakeholder alpha are important in this process. In this sense, some of the practices in this category can be related to other practices under other alphas. Progress on business model development can also be tracked by evaluating how well the current business model is functioning and to what extent it is already operational.

Thirdly, marketing may warrant its own alpha, should the Essence be expanded to include business aspects. Marketing is as important to startups as it is to any other type of company [3]. In fact, as startups generally have less capital to spend on marketing, they are forced to spend more time coming up with more creative marketing approaches. While various marketing practices arguably exist, this potential alpha also overlaps with the stakeholder alpha. Marketing is focused on users and customers and they are already at the center of the stakeholder category in Essence, when used by a startup. However, marketing practices that are not user-centric would be difficult to incorporate into the stakeholder category.

Alternatively, one other option would be to look at other theories and frameworks commonly utilized by startups for business model development. Potential business-related alphas could be derived e.g. from the Business Model Canvas [17].

## 5.1    Practical Implications

The primary practical contribution of this study are the practices listed in the tables in the results section, combined from the results of Dande et al. [5] and our data analysis. These practices can help guide work in software startups. Moreover, they can be used to construct methods in conjunction with other practices.

Additionally, based on some of the practices and the interviews, we can draw some further practical implications:

- Flat organization and self-organizing teams seem to be an effective way for constructing the initial team. Self-organizing teams have been noted to be beneficial in Agile [10]. It may also be beneficial to avoid strict roles.
- You should have a clear idea of what is the core product and what features are the key features at any given moment. Having a scope too large for the product or an MVP is a frequent reason for failure in software startups [12].
- Forming close relationships with initial customers and users is beneficial. They can help you develop your product and participate in development. They can also aid in marketing. For example, user communities on social media platforms built around your (future) product can be beneficial in various ways.

## 5.2    Limitations of the Study

One key limitation of this study is how we have chosen to define a 'common' practice. We included practices that were discussed by two or more case startups (n=13), considering them sufficiently common. As this approach was qualitative, we cannot make any claims as to how common these practices really are among startups. Thus, we opted to rather include too many practices than too few practices, so that future studies might investigate how frequently they really are utilized and exclude any uncommon ones. Indeed, not all practices listed by Dande et al. [5] were discussed by the case startups either, further highlighting this limitation.

Moreover, the qualitative case study approach in general can always be considered an issue in terms of generalizing the results. However, we argue that 13 cases is a large enough number for some generalizability. For example, Eisenhardt [6] suggests that a qualitative case study into a novel topic have at least five cases.

# 6 Conclusions

In this paper, we have studied Software Engineering (SE) in software startups from the point of view of commonly used practices. Data were collected through semi-structured interviews in a case study of 13 software startups. This data set was used to complement the results of an existing study that produced a list of 63 practices [5], confirming some of their results while uncovering some new practices.

We studied the suitability of the Essence Theory of Software Engineering for the startup context. Our results suggest that the software startup context involves business aspects in ways a generic SE project does not. Thus, the theory fails to account for many of the more business-related practices utilized by the case startups. We propose that the theory either be expanded to create a more startup-oriented version with business alphas, or that other theories or tools (such as the Business Model Canvas [17]) are used in conjunction to cover the business aspects.

Additionally, we provide an extensive list of 76 software startup practices for use in method engineering for software startups. 63 of the practices are those proposed by Dande et al. [5] while 13 are ones proposed based on the empirical data here.

Future studies should particularly look into extending the Essence Theory of Software Engineering for software startup use. The alphas we have proposed here can be utilized for this purpose. However, they need to be formalized to fit the theory.

## References

1. Blank, S.: The four steps to the epiphany: successful strategies for products that win. BookBaby (2007).
2. Chang, S. J.: Venture capital financing, strategic alliances, and the initial public offerings of Internet startups. Journal of Business Venturing, 19(5), 721-741, (2004).
3. Crowne, M.: Why software product startups fail and what to do about it. Evolution of software product development in startup companies. In Proceedings of the 2002 Engineering Management Conference IEMC'02, pp. 338-343, IEEE (2002).
4. Cruzes, D. S., Dyba, T.: Recommended steps for thematic synthesis in software engineering. In Proceedings of the 2011 Symposium on Empirical Software Engineering and Measurement (ESEM), pp. 275-284, IEEE (2011).
5. Dande, A., Eloranta, V. P., Kovalainen, A. J., Lehtonen, T., Leppänen, M., Salmimaa, T., ... Koskimies, K.: Software startup patterns - an empirical study. Tampereen teknillinen yliopisto. Tietotekniikan laitos. Raportti-Tampere University of Technology. Department of Pervasive Computing. Report; 4 (2014).
6. Eisenhardt, K. M.: Building theories from case study research. Academy of Management Review, 14(4), 532-550 (1989).
7. Giardino, C., Paternoster, N., Unterkalmsteiner, M., Gorschek, T., Abrahamsson, P.: Soft-ware Development in Startup Companies: The Greenfield Startup Model. IEEE Transactions on Software Engineering, 42(6), 585-604 (2016).
8. Jacobson, I., Ng, P. W., McMahon, P., Spence, I., Lidman, S.: The essence of software engineering: the SEMAT kernel. ACM Queue, 10(10), 40 (2012).
9. Kamulegeya, G., Hebig, R., Hammouda, I., Chaudron, M., Mugwanya, R.: Exploring the Applicability of Software Startup Patterns in the Ugandan Context. In Proceedings of the 43rd Euromicro Conference on Software Engineering and Advanced Applications (SEAA), pp. 116-124, IEEE, (2017).
10. Karhatsu, H., Ikonen, M., Kettunen, P., Fagerholm, F., Abrahamsson, P.: Building Blocks for Self-Organizing Software Development Teams a Framework Model and Empirical Pilot Study. In Proceedings of the 2nd International Conference on Software Technology and Engineering (ICSTE) (2010).
11. Klotins, E.: Software start-ups through an empirical lens: are start-ups snowflakes? In Pro-ceedings of the International Workshop on Software-intensive Business: Start-ups, Ecosys-tems and Platforms (SiBW) (2018).
12. Klotins, E., Unterkalmsteiner, M., Gorschek, T.: Software Engineering Antipatterns in start-ups. IEEE Software, 36(2), 118-126, (2018).
13. Klotins, E., Unterkalmsteiner, M., Gorschek, T.: Software engineering in start-up companies: An analysis of 88 experience reports. Empirical Software Engineering, 24(1), 68-102. (2019)
14. Lueg, R., Malinauskaite, L., Marinova, I.: The vital role of business processes for a business model: the case of a startup company. Problems and Perspectives in Management, (12, Iss. 4 (contin.)), 213-220, (2014).

15. Melegati, J., Goldman, A., Paulo, S.: Requirements Engineering in Software Startups: a Grounded Theory approach. 2nd Int. Work. Softw. Startups, Trondheim, Norw. (2016).

16. Nguyen-Duc, A., Wang, X., Abrahamsson, P.: What Influences the Speed of Prototyping? An Empirical Investigation of Twenty Software Startups. In Proceedings of the 2017 International Conference on Agile Software Development (XP2017), pp. 20-36. (2017).

17. Osterwalder, A., Pigneur, Y., Clark, T.: Business Model Generation: A Handbook for Vi-sionaries, Game Changers, and Challengers. Hoboken, NJ: Wiley (2010).

18. Paternoster, N., Giardino, C., Unterkalmsteiner, M., Gorschek, T., Abrahamsson, P.: Soft-ware development in startup companies: A systematic mapping study. Information and Software Technology, 56(10), 1200-1218 (2014).

19. Tolvanen, J. P.: Incremental method engineering with modeling tools: theoretical principles and empirical evidence. Ph. D. Thesis, University of Jyvaskyla (1998).

20. Unterkalmsteiner et al.: Software Startups - A Research Agenda. E-Informatica Software Engineering Journal, 1, 89-124 (2016).